\documentclass[12pt,a4paper]{article}
\usepackage{pazh}
\usepackage[cp1251]{inputenc}
\usepackage[english]{babel}
\usepackage{graphicx}
\DeclareGraphicsExtensions{.pdf,.png,.jpg}
\usepackage{indentfirst}
\usepackage{afterpage}
\usepackage{longtable}
\usepackage{amsmath}
\usepackage{amssymb}

\def\gee{ \, \lower 1mm\hbox{$\,{\buildrel > \over{\scriptstyle\scriptstyle\sim} }\displaystyle \,$}}
\def\lee{ \, \lower 1mm\hbox{$\,{\buildrel < \over{\scriptstyle\scriptstyle\sim} }\displaystyle \,$}}
\def\|{\partial}

\def\varkappa {{\scriptstyle\partial}\! e}

\let\c=\centerline

\let\b=\baselineskip
\begin{document}
\headheight 1.50true cm \headsep  0.7true cm
\righthyphenmin=2


\large


\begin{center}
 \textbf{HI content in the galactic discs: the role of gravitational instability.}
\end{center}

\c {Zasov A.V.$^{(1,2)}$, Zaitseva N.A.$^{(1)}$}

\c {\it \scriptsize $^{(1)}$Sternberg Astronomical Institute, Moscow State University, Moscow, Universitetskii pr. 13, Russia}

\c {\it \scriptsize $^{(2)}$Department of physics, Moscow State University, Moscow, Universitetskii pr. 13, Russia}

\bigskip
\textbf{ABSTRACT}

We examine the dependence between hydrogen total mass $M_{HI}$ and rotation speed $V_{rot}$, optical size $D_{25}$ or disc radial scale $R_0$ for two samples of late-type galaxies: a) isolated galaxy ($AMIGA$ sample), and b) the edge-on galaxies (flat galaxies of Karachentsev et al. 1999). Estimates of $M_{HI}$, given in the $HYPERLEDA$ database for flat galaxies appear to be on average higher at $\sim $0.2 dex, than for isolated galaxies with similar $V_{rot}$ or $D_{25}$ values, most probably, due to the overvaluation of self-absorption in the $HI$ line. We confirm that the hydrogen mass for both samples closely correlates with galactic disc integral specific angular momentum $J$, which is proportional to $V_{rot}D_{25}$ or $V_{rot}R_0$, with low surface brightness galaxies lie along a common $V_{rot}R_0$ sequence. This relationship can be explained, assuming that gas mass in the disc is regulated by marginal gravitational stability condition of gas layer. A comparison of the observed and theoretically expected dependences leads to a conclusion that either  gravitational stability corresponds to higher values of $Toomre$ parameter than is usually assumed, or the threshold stability condition for most galaxies took place only in the past, when gas mass in discs was 2-4 times higher than at present (with the exception of galaxies with abnormally high $HI$ content). The last condition requires that the gas accretion was not compensated  by gas consumption during the evolution of most of galaxies.

\medskip
\c{\bf 1. Introduction}
 \medskip

Discy galaxies  vary greatly in cold gas content and, as a consequence, in starformation rate and related characteristics, because there is a large number of factors influencing gas flow to a disc and its consumption or outflow (see Fig. 1). The main gas loss channels are the turning gas into stars, as well as  blowing gas out from a disc to halo or intergalactic space due to AGN or young stars activities (supernovae and stellar winds). In turn, gas supply is provided by gas losses by evolved stars, by cooling of hot halo gas and its accretion (hot mode accretion), by accretion of intergalactic gas filaments (cold mode accretion), as well as by destruction and absorption of dwarf satellites containing interstellar medium. Gas may also be ``blown out'' of galaxy disc due to its interaction with the environment.

Galaxies  reveals a bi-modal distribution if to consider their gas content or starformation rates ($SFR$) per mass unit of stellar population (see., e.g., Brammer et al. 2009). Conventionally, galaxies are divided into passive (quiescent) galaxies with low gas content and little or no young stars, belonging mostly to $E-S0-Sa$ types, and  starforming ones, where gas mass is large enough to maintain starformation. Most of actively starforming galaxies belong to $Sbc$ and later types. These are the galaxies discussed in this paper. 

Gas losses may transfer a galaxy from starforming into passive category, and the transition time must be short enough to explain  a bi-modal distribution of their color or $SFR$ (see, for example, Bundy et al. 2010; Volcani et al. 2015). But, as long as galaxies remain to be starforming ones, their evolution proceeds surprisingly similar. This is evidenced by the existence of correlations between neutral hydrogen integrated mass $M_{HI}$ and such global slowly evolving parameters, as the optical or gaseous disc radius (Solanes et al. 1996; Karachentsev et al. 2004; Toribio et al. 2011; Wang et al. 2016), stellar disc luminosity or mass (Karachentsev et al. 1999; Bradford et al. 2015; Lell et al. 2016; Toribio et al. 2011; Evoli et al. 2011), galaxy rotational velocity (Begum et al. 2008; Toribio et al. 2011), halo mass or spin (Evoli et al. 2011; Huang et al. 2012), and disc specific rotation angular momentum $J$, which is proportional to the product of rotational velocity  by disc optical radius (Safonova 2011; Zasov, Sulentik 1994; Zasov, Smirnova 2005; Obreschkow et al. 2016).

It is evident that the above correlations may not be independent, although a priori it is not clear which of them are physically conditioned, and which just reflect a mediated relationship between various galaxies parameters. For example, there exists a relationship between disc rotation speed and a size of a galaxy, which is close to linear (see., eg., Karachentsev et al. 2013), so the relations between $M_{HI}$ and disc size and disc specific angular momentum appear to have a common cause.

The most tight relationship is observed when $M_{HI}$ is compared with radius $R_{HI}$ which corresponds to the fixed azimuthally averaged gas surface density, usually taken as $\Sigma_{HI} \approx 1 M_\odot/ pc^2$ (Lelli et al. 2016; Wang et al. 2016). However, these matching parameters cannot be considered as independent ones: a change of $M_{HI}$ during the evolution will inevitably lead to a shift of $R_{HI}$. Nevertheless, the close correlation between $M_{HI}$ and $R_{HI}$, which takes place for galaxies with different masses -- from low-mass gas rich dwarf to giant spirals -- indicates that the shape of the overall $HI$ radial profile should be similar for most galaxies: only in this case $R_{HI}$ will be associated with hydrogen total mass $M_{HI}$. 
Indeed, radial profiles $\Sigma_{HI}(R)$ for late-type galaxies demonstrate their universal character except the central regions of galaxies, where $HI$ distribution is very diverse, partially because a substantial, if not most of gas there is usually molecular (Bigiel, Blitz 2012; Martinsson et al. 2016). They look particularly similar in shape within a large range of $R$, if the radial distance is expressed in units of radius $R_{HI}$  (Swaters et al. 2002; Wang et al. 2016) or is normalized to optical radius $R_{25} = D_{25}/2$, which corresponds to the boundary isophote 25 magnitude/sq.sec. (Bigiel, Blitz 2012). Such similarities, in turn, suggest a similar character of gas evolution for galaxies with different masses.

The mean observed profile $\Sigma _{HI}(R/R_{25})$, obtained by Bigiel al. (2010), evidences that about half of $HI$ integral mass contains inside the optical radius $R_{25}$. However, for dwarf galaxies, this fraction may be much lower. According to Martinsson et al. (2016) the average $R_{HI}/R_{25}$ for spiral galaxies is $1.35\pm 0.22$.

It's worth noting that the contribution of $H_2$ in the total gas mass is usually small. According to Bothwell et al. (2014), $H_2$ mass fraction in spiral galaxies averages 0.09-0.13 and decreases parallel with mass of stellar population. Lisenfeld et al. (2011) obtained the average ratio $M_{H_2}/M_{HI} \approx 0.2$ for isolated galaxies of $AMIGA$ sample, used in this paper. Thus,  a total mass of atomic $HI$ may characterize a total neutral gas mass.

In this paper we examine the empirical relationships between $M_{HI}$ and optical radius, disc radial scale and specific angular momentum. Also we analyze the possible relationship between the observed $HI$ mass and the marginal (threshold) stability condition of gas layer to the gravitational perturbations in a disc plane. 

\medskip
\c{\bf 2. Relation between hydrogen mass and disc size and rotation.}
 \medskip

We used two  samples of late-type galaxies: the Isolated Galaxies Catalogue $AMIGA$ (Lisenfeld et al. 2007) created on the basis of the Karachentseva Isolated Galaxies Catalogue (Karachentseva et al. 1986) (we consider the objects with disc inclination $i>35^0$) and flat galaxies sample, containing edge-on galaxies with a large axis ratio $(a/b)$ from the catalogue $RFGC$ by Karachentsev et al. (1999). The latter galaxies represent most homogeneous late-type galaxies sample which possess a similar structure (a thin disc and a small bulge). We excluded from consideration the galaxies, if the errors of rotational velocity (taken from $All\, Digital\, HI$ directory Courtois et al. 2009) exceed 20 km/s, as well as galaxies with angular size larger than 400" (to achieve the better homogeneity of data, as much as the estimation of total $M_{HI}$ is more complicated and less accurate for galaxies with large angular size). Both samples do not contain clearly interacting galaxies. Gas mass $M_{HI}$, linear optical sizes $D_{25}$, inclination (for $AMIGA$ galaxies) and rotation speed $V_{rot}$ were taken from the $HYPERLEDA$ database (leda.univ-lyon1.fr, Makarov et al. 2014) or were found from the data contained therein. Disc radial scales lengths $R_0$ are based on $SDSS$ review photometry images in the $i$-band, and were taken from Hall et al. (2012). The accepted distance scale corresponds to $H_0=75$ km/s/Mpc.

Fig. 2 illustrates three-dimensional relationship $M_{HI}-V_{rot}-D_{25}$ for isolated galaxies. Flat galaxies give a very similar pattern (not shown here), although, as it will be discussed below, there is a notable shift between flat and isolated galaxies. Regression parameters $k$ and $b$, the standard deviations $MSE$ and correlation coefficients $p$ are given in Table 1. The slope and zero point of these regression lines were calculated for a bisector line between direct and inverse regressions. We tried two ways for regression construction: a simple LSM and a robust regressions. The most deviating points are attributed to the lower statistical weights in the last option. However, in both cases we obtained practically identical results.

For both samples of galaxies the most close correlation is between $M_{HI}$ and a disc size or specific angular momentum. Although $M_{HI}$ correlation with $V_{rot}$ is not so tight, it may not be a simple reflection of well-known dependence ``optical size vs rotation speed'' (see, for example, Russell 2002), as it is evidenced by the existence of correlation between $V_{rot}$ and the deviation from the straight line $logM_{HI}- logD_{25}$  for $AMIGA$ sample (Fig. 3). 

The situation is more complex for flat galaxies. Their $M_{HI}$ values are systematically higher than for isolated galaxies for all relations we consider (at least for galaxies with large enough $D_{25}$ or $V_{rot}$). For example, in Fig. 4a, b flat galaxies and $AMIGA$ isolated galaxies (presented by the regression line) are compared at ``$M_{HI}-V_{rot}$'' and ``$M_{HI}-D_{25}$'' diagrams. Apparently, this shift is a result of $HI$ mass overestimation for flat galaxies due to the unreliability of accounting of large self-absorption in $HI$ line. Flux extinction  caused by internal absorption is less than 40\% for inclination $i=85^0$, however it increases rapidly as we approach to $i \approx 90^0$ (Heidmann et al. 1972). In $HYPERLEDA$ database, self-absorption correction for edge-on galaxies was assumed to be (expressed in stellar magnitudes) $\Delta m_{HI}$ = -0.82, which corresponds to the attenuation coefficient $k$ = 2.1. The ``shift'' between flat and isolated galaxies (there are only a few edge-on galaxies among the latter) demonstrated in the diagrams allows to propose that the  self-absorption is overrated and hence $M_{HI}$ is overestimated by about 1.5 times on average. The naturally expected scatter of the attenuation coefficients for different edge-on galaxies, of course, increases the points dispersion on the diagrams. Curiously, a difference between $logM_{HI}$ estimates for edge-on and isolated galaxies becomes insignificant, or even change its sign for slowly rotating, and, consequently, low-mass galaxies ($logV_{rot}<$1.9, or $V_{rot}<$80km/s) as it follows from Fig. 4b. Although the number of such galaxies is too small for reliable conclusions, one may suggests that the accepted  self-absorption in $HI$ line for these galaxies is not overestimated  at all, apparently because the $HI$ line-of-sight velocities have lower dispersion there, which makes self-absorption more significant.

\medskip
\c{\bf 3. Gas content and gravitational stability of gas layer.}
 \medskip

As it was noted in the Introduction, the correlation between the current hydrogen mass and the slowly evolving characteristics of galaxies (size, rotational speed, specific angular momentum) indicates similar character of gas evolution for the bulk of starforming galaxies. Note however that the disc size of galaxies is still subject to slow evolution because their stellar population and hence the brightness distribution changes over time. A photometry of late-type galaxies without noticeable bulges in the Hubble Deep Field ($HDF$) (Sachdeva, Saha 2016) demonstrates a slow increase of disc diameter from $z$ = 1 reaching several tens of percent. However, this does not apply to the discs radial scales $R_0$ which do not reveal a significant change during this time interval. Therefore $R_0$ can be considered as more conservative parameter, evolving more slowly than the isophotal diameter.

Rotational velocity of a galaxy, established after its disc was formed, is also a conservative value, which can change as the result of large-scale mass redistribution only, caused, for example, by strong interaction with neighbor galaxies. Therefore, the correlation between gas mass and such parameters as $D_{25}$, $R_0$ or $J\sim V_{rot}R_0$ evidences that either a gas content of galaxies remains almost unchanged in the last few billion years (it requires the accretion to compensate the process of gas losses), or most galaxies evolve in a similar way, so that the sequences at the diagrams  have not ``blurred'' over billions of years.

In principle, both options are possible. Numerical models were proposed to describe the galaxy evolution, where star formation and accretion compensate each other, and, as a result, the integrated gas mass maintains at nearly constant level (so called ``$bathtub$'' models; see, for example, Strinson et al. 2015). However, in this case the important questions remain. What defines the equilibrium level for surface density $\Sigma_{HI}$ or total mass $M_{HI}$ in galaxies? Why should this level depend on the optical size or the angular momentum of a disc, instead of for example, on the environment density? Why similar gas density distribution $\Sigma_{HI}(R)$ is maintained for a significant part of galactic discs? One would rather expect that the accretion rate and the resulting gas distribution inside a disc should be different for galaxies with different masses and morphological types, as well as for isolated galaxies and those in a dense environment.

From another point of view, the condition of local gravitational stability of rotating gas layer can serve as the control of gas evolution. Indeed, a development of instability should lead to the increase of local density inhomogeneities in a disc at kpc-scale, and, as a consequence, to the intensification of starformation till the threshold (marginally stable) state of rotating gas layer is reached.

The idea that the gas layers of galaxies are in many cases close to the marginally stable state was proposed to explain the observed distribution of gas in galaxies by many authors, beginning with the papers by Quirk (1972), Zasov and Simakov (1988), Kennicutt (1989), Martin and Kennicutt (2001). The indirect arguments for gravitational stability role in the gas evolution were later presented by Zasov and Terekhova (2013) and Meurer et al. (2013). These authors showed that the observed correlation between radial profile $\Sigma_{HI}(R)$ and radial distribution of dark halo column density in galaxies, found from the rotation curve modeling, may be explained, if to assume that gas density profiles $\Sigma_g (R)$ follow the marginal stability condition.

However, the assumption of marginal stability of gas layer in a sufficiently extended range of $R$ encounters difficulties when applied to the outer regions of spiral galaxies, as well as to $Irr$-galaxies. The analysis of observational data shows that a gas layer usually has a much lower density than it is required for the threshold (marginal) stability condition (see, e.g., Kim, Ostriker 2007; Westfall et al. 2014; Elmegreen, Hunter 2015).

Below we consider this issue in more detail.

The stability condition against local gravitational perturbations is usually characterized by the $Toomre$ parameter: 
\begin{equation}\label{Eq1}
Q_g=\frac {c_g\varkappa}{\pi G \Sigma_g}, 
\end{equation}
where $c_g$ is a gas velocity dispersion associated with turbulent motions, and $\varkappa$ -- the epicyclic frequency. For the constant circular velocity $V_c$, $\varkappa = \sqrt{2}V_c/R$. Here we assume that the gas circular velocity is close to the rotational speed found for $HI$ layer (that is $V_c \approx V_{rot}$).

In a simplest case of a thin axisymmetric disc the critical value $Q_{c,g}$ for stability parameter $Q_g$ against radial perturbations is $Q_{c,g}=1$. The accounting of non-radial perturbations make a disc less stable, while the non-zero thickness stabilizes it. When the marginal stability is considered, these two factors to a large extent obviate each other. The analysis of stability conditions for real galaxies is simplified by the slow variation of gas one-dimensional dispersion $c_g$ along the radius and its similarity for galaxies with different luminosity (mass) with $c_g\approx $ 6-10 km/s for the outer regions (see the discussion in Leroy et al. 2008; Ianjamasimanana et al. 2015).

It should be noted that all analytically derived criteria of disc (in)stability are approximate and local, and in general, a construction of three-dimensional numerical models is required for analysis of gravitational perturbations growth. Such models show that critical values of $Q_{c,g}\sim $1.2-2 depending on the radial distance $R$ and the ratio between a disc mass and a mass of spheroidal subsystems within a given $R$ (see, for example, Khoperskov et al. 2003; Kim, Ostriker 2001; Kim, Ostriker 2007). In the numerical 3D-models (Kim, Ostriker 2007) of the disc, which parameters are close to those observed for near-solar neighborhood, a critical stability parameter for gas layer was found to be $Q_{c,g}$ = 1.4. The analytical $Q_{c,g}$ assessment which takes into account non-axisymmetric disturbances leads to a similar conclusion. For gaseous disc with flat rotation curve Poliachenko et al. (1997) obtained $Q_{c,g}= \sqrt{3}$.

If we use the most probable values of $c_g$ and $Q_{c,g}$, and assume that gas density corresponds to the gas layer marginal stability (Eq. 1), then the mass of gas within a given radius $R_{lim}$ may be found from the equation
\begin{equation}\label{Eq2}
M^c _{gas}= \int \limits_0^{R_{lim}} 2\pi R\Sigma_g(R)\,dR \,=\,2\frac {c_g}{Q_{c,g}}\int \limits_0^{R_{lim}}R\varkappa dR,
\end{equation}
which shows that a gas mass is determined by radial changes of $\varkappa (R)$.

Let the rotation curve has a form $V(R)=V_c({R}/{R_{lim}})^n$, where $V_c$ is circular velocity at $R=R_{lim}$ ($n=0$ for flat rotation curve and $n=1$ for linearly increasing speed). Then $\varkappa (R)= \sqrt{2}\Omega (R)(1+n)^{1/2}$, where angular velocity $\Omega (R) = V(R)/R$. For the critical value $\Sigma_g = \frac {c_g}{Q_{c,g}}\frac {\varkappa}{\pi G}$ a total gas mass within $R_{lim}$ will be:

$M^c_{gas} =2^{3/2}K(1+n)^{-1/2}V_cR_{lim}$, 

where the factor $K = (c_g/Q_{c,g})$ is considered to be approximately constant along the radius. This implies that a total mass of marginally stable gas layer weakly depends on rotation curve shape, being proportional to the disc specific angular momentum $J\sim V_cR_{lim}$. For a flat rotation curve ($V_c = const$) the atomic gas mass $M^c_{HI}=2^{3/2}\eta \frac {K}{G}\cdot V_cR_{lim}$, where $\eta^{-1}\approx $0.5-0.7 is a transition coefficient from $M_{gas}$ to $M_{HI}$, which takes into account a fraction of molecular hydrogen and helium in the total gas mass.

In spite of the approximate nature of this approach, it nicely agrees with observations, showing a direct proportionality between $M_{HI}$ and specific angular momentum $J$ within the optical radius $R_{lim}=R_{25}$ for single galaxies and group members (Karachentsev et al. 2004; Safonova 2011; Zasov, Rubtsova 1989; Zasov, Sulentic 1994; Zasov, Smirnova 2005). Given that the total galaxy mass within $R_{25}$ is $M_{tot}\approx \frac {V_c^2R_{25}}{G}$, a gas mass over galaxy mass ratio is: 
\begin{equation}\label{Eq3}
F\equiv \frac {M_{gas}}{M_{tot}} \sim \frac {c_gJ}{Q_{c,g} \cdot V_c^2R_{25}}\,\sim \frac {c_g}{Q_{c,g}V_c},
\end{equation}
i.e., the relative gas mass is lower for fast rotating galaxies, that is actually takes place for spiral and irregular galaxies (see., eg., Fig. 14 in the catalog $UNGC$ by Karachentseva et al. 2013). 

It should be borne in mind that this simple form of dependence between gas mass and specific angular momentum has the approximate character due to a number of simplifications, the most important of which is the assumption of $c_g/Q_g$ constancy. This condition is hardly acceptable for the inner region of discs, where the observed gas density radial profiles are very diverse, and also for far peripheral areas (outside optical boundaries generally), where a significant increase in the gas layer thickness plays a stabilizing role, increasing the stability parameter $Q_{c,g}$.


In Figs. 5 a,b we presented the relationship between $M_{HI}$ and $J$ (to be more precise -- between  the total gas mass assumed to be proportional to $M_{HI}$ and  $V_{rot}D_{25}$ (a) or $V_{rot}R_0$ (b) parameters) for $AMIGA$ galaxies. The solid straight line at both diagrams is a linear regression for these galaxies, and two parallel dashed lines represent the expected relations for marginally stable discs with $\eta = \frac {M_{HI}}{M_{gas}}$= 0.5 for two probable values of $K$ = $(c_g/Q_{c,g})$ = 10 km/s (upper line) and 5 km/s (bottom line). $HI$ mass within $D_{25}$ was taken equal to a half of the integral atomic hydrogen mass, as it is performed for spiral galaxies in average (Bigiel, Blitz 2012). Upper dashed line corresponds to the stability parameter  $Q_{c,g}$ = 1, and the lower one -- to $Q_{c,g}$ = 2  for gas velocity dispersion $c_g$ = 10 km/s. Points scatter on the graphs mainly reflects the difference of fraction of $M_{HI}$ enclosed within $R_{25}$ for different galaxies. As it follows from Figs. 5 a,b, most of galaxies lay below the strip bounded by the dashed lines for high and low $K$-values. This shift opens the question whether gas layers of galaxies are really in most cases close to marginal stability, or they pass a significant stability reserve ($Q_g>$2). 

Regardless of the answer, these diagrams can be used for $HI$ content diagnostic, allowing to reveal the abnormally high, or, conversely, abnormally low gas mass in comparison with isolated galaxies with similar kinematic parameters. As an example, we compare the $AMIGA$ galaxies with galaxies  unusually rich of  $HI$ from the $HIghMass\, galaxy\,sample$ (Huang et al. 2014). In these gas-rich galaxies $HI$ mass exceeds $10^{10}M_{\odot }$, reaching in most cases of more than one-third of the total mass of stellar population. These galaxies are marked with diamonds at Figs. 5 a,b. Their $HI$ masses are systematically higher than for the other galaxies with the same angular momentum, revealing the peculiar way of gas evolution, possibly related to the unusually low efficiency of starformation or accretion. 

In Fig. 5 b we also compare $AMIGA$ galaxies with low surface brightness galaxies ($LSB$), considered earlier by Abramova and Zasov (2011) (marked with asterisks); we  also included the unusual dwarf galaxy, $LSB$-companion of spiral galaxy $NGC4656$ first described by Schechtman-Rook, Hess (2012). This galaxy has an extremely low surface brightness with the enhanced glow in ultraviolet, which indicates the current or recent starformation. Here we use a disc radial scale $R_0$ instead of isophote diameter $D_{25}$ to avoid the dependence on a disc surface brightness. $M_{HI}$ and $V_{rot}$ values were taken from the paper of Abramova and Zasov (2011), the data for $NGC4656$ are used from Schechtman-Rook, Hess (2012) and Zasov  et al. (2017). One can see that $LSB$-galaxies follow general relationship with  normal brightness galaxies, although with a larger scatter. Some $LSB$-galaxies with high specific momentum have abnormally high $M_{HI}$ in comparison with $AMIGA$ galaxies, being in the same place as the galaxies with abnormally high $HI$ mass (diamonds). Low accuracy of parameters and poor statistics do not allow to make reliable conclusions from this comparison, however it  suggests a similar regulatory factors of gas evolution for $LSB$ and normal brightness galaxies.

Now we proceed from the total gas mass to the Toomre' parameter $Q_g$ at different radial distances, in order to clarify how far a gas layer is from the stability threshold condition. Direct estimates (Leroy et al. 2008; Bigiel et al. 2010; Romeo and Falstad 2013; Westfall et al. 2014; Obreschkow et al. 2016; Yim et al. 2011) showed that in the outer regions of galaxies $Q_g$, determined by the classical expression (1) for the azimuthally averaged gas density, reaches several units (generally, $Q_g \sim $2 - 4 and higher outside $R_{25}$). It confirms that  gas layers are in general stable, if the density critical value corresponds to $Q_g = Q_{c,g}\approx$ 1, although gas density may be close to critical one in the inner part of a galaxy, and have a significant stability reserve at a periphery. 

The idea that at some radial distance gas surface density passes through the stability threshold, so starformation rate is low outside this zone, was first considered (taking into account molecular gas) by Martin and Kennicutt (2001). Authors concluded that the stability threshold usually takes place at radial distance where the ratio $K=\frac {c_g}{Q_{c,g}}$ falls below 4 km/s. It corresponds to $Q_{c,g}\approx 1.5-2.5$ for realistic values $c_g\sim $6-10 km/s at a disc periphery. However, as it follows from the radial distribution of UV brightness, starformation  often continues without experiencing a sharp drop where the transition to a stable mode is expected (Boissier et al. 2007). It evidences that a formation of stars is a local process, which is determined by local fluctuations of gas density and temperature (see the discussion in Elmegreen 2011), while the gravitational instability of rotating disc develops at scales around a Jeans wavelength  $\lambda_J$, which is $\sim$1 kpc and even larger for massive galaxies. 

The important factor which lowers a stability threshold allover a galactic disc is a destabilizing role of stellar disc gravity. A good approximation for stability criterion of two-component disc $Q_{sg}$ is the sum of reciprocals of $Q$-parameters taken separately for gas ($Q_g$) and stars ($Q_s$) with a weighting factor of $W$, which reduces the contribution of component with higher $Q$ (see Romeo, Wiegert 2011; Romeo, Falstad 2013).
\begin{equation}\label{Eq4}
\frac {1}{Q_{sg}}=\begin{cases}
\frac {W}{Q_s}+\frac {1}{Q_g},&\text{if $Q_s\ge Q_g$}\\
\frac {1}{Q_s}+\frac {W}{Q_g},&\text{if $Q_g\ge Q_s$}\\
\end{cases}
\end{equation}
\begin{equation}\label{Eq5}
W = \frac{2c_sc_g}{c_g^2+c_s^2}
\end{equation}

According Leroy et al. (2008), stellar and gas components give a comparable contribution to  general disc stability. Parameter $Q_g$ for gas component is around 2-3, being several times higher at the far periphery of discs. At the same time, the combined gas-stellar parameter $Q_{sg}\approx 2$ within a wide range of $R$. Later Meurer et al. (2013) analyzed the radial profiles of $Q_g(R)$ and $Q_{sg}(R)$ for the sample of 20 galaxies (late types spiral galaxies from $THINGS$ review mostly) with known $HI$ density profiles. They showed that on average about 50\% of total $HI$ mass lies within the range of radial distances where $\Sigma_{HI}(R)$ decreases approximately as $1/R$ and correlates with the epicyclic frequency profile $\varkappa (R)$ in such a way, that $Q_g$ and $Q_{sg}$ parameters remain approximately constant for the fixed $c_g$, although both of them exceed unit. In turn, Romeo and Falstad (2013) concluded for $THINGS$ galaxies, that gas plays a dominant role in stability of the inner parts of discs (where a significant part of gas is molecular one), while stellar disc determines a stability parameter at large distances from the center.

However, in the papers cited above a relative role of stellar disc seems to be exaggerated,  especially at large radial distances. Firstly, in spiral and irregular galaxies gas velocity dispersion $c_g(R)$ slowly decreases with $R$, falling below 10 km/s at $R=(1-2)R_{25}$ (Leroy et al. 2008; Walter et al.  2008; Ianjamasimanana et al. 2015). Secondly, the indirect estimate of stellar velocity dispersion is usually obtained from a disc half-thickness (a vertical scale), which is regarded as independent of radius (Leroy et al. 2008; Romeo, Falstad 2013), whereas in real galaxies it increases with $R$ in the outer regions (de Grijs, Peletier 1997; Yim et al. 2014), which reduces the destabilizing role of stellar disc. For example, the increase of disc thickness at a factor of 2  leads to an increase of stellar velocity dispersion estimate for about $\sqrt{2}$ times. At the same time, the weighting coefficient $W$ in Eq. 4-5 is reduced at about the same factor, so that the relative contribution of stellar disc in the resulting value $\frac {1}{Q_{sg}}$ (see Eq. 4)  will be approximately halved. As a result, gas component may play a more  significant input in stability parameter. Nevertheless, the conclusion about gravitational stability at large $R$ will hardly changed. 

As the illustration, in Table. 1 we present the values of parameter $Q_g$ at $R=R_{25}$ for  $THINGS$ galaxies with known rotation curves (de Blok et al., 2008) and gas surface density profiles $\Sigma_g=1.3\Sigma_{HI}$ (Walter et al. 2008). The direct estimates of $HI$ velocity dispersion were taken from Ianjamasimanana et al. (2015). As one can see, in most cases $Q_g$ at the edge of optical discs ranges between 3 and 6, which confirms a significant stability reserve of gas layer.

\medskip
\c{\bf 4. Threshold stability of gas layer or synchronous evolution?}
 \medskip

Now we are faced with a contradictory situation. On the one hand, the shape of $HI$ radial profile of surface density in galaxies in a wide range of $R$ is close to the anticipated one for $Q_g=const$. This is confirmed by correlation between the gas mass in galaxies and disc kinematic parameters. On the other hand, the existing estimates show that $Q_g$ exceeds the expected critical value $Q_{c,g}\approx 1.5-2$, which is required for marginal stability. Primarily, it refers to the outer regions of galaxies, where $Q_g$ reaches 5 or higher. A stellar disc gravity may bring a disc closer to the stability threshold ($Q_{sg}\lesssim 2$). However, in this case it becomes not clear, how to explain the  observed consistency between gas mass $M_{HI}$ or density $\Sigma_{HI}$ and the  angular momentum, expected for marginally stable gas layer?

We consider two possible options.

\textit{\textbf{Option 1.}} One can propose that  for a significant fraction of atomic gas the velocity dispersion is 2-4 times lower than the commonly used value (about 10 km/s). However, observations do not provide sufficient grounds for such conclusion. As an alternative, it can be assumed that gas layer stability threshold corresponds to a higher values $Q_{c,g} \approx $3-4, for example, as a result of energy dissipation, making gas layer less resistant to gravitational perturbations (see, eg., Elmegreen 2011). Another factor which may explain too high  azimuthally average values of $Q_g$ is the inhomogeneous gas distribution. Even if the average value of $Q_g$ obviously evidences the stability, some fraction of gas may be located in the areas where $\Sigma_{HI}$ is several times higher than the mean value at a given $R$, being close to the stability threshold.  However, it cannot explain why the shape of azimuthally averaged profile $\Sigma_{HI}(R)$ in this case should follow the equation (1) expected for $Q_{g}=const$.

\textit{\textbf{Option 2.}} Gas layers of galaxies could reach marginally stable state several billions years ago, when gas density and the total gas mass were at least twice as high as at present time, so the close relationship between specific angular momentum and $M_{HI}$ was established that time. This could be the case,  when  a formation of stellar disc was almost finished, and gas turbulent velocity, high at the violent disc formation stage, decreased to the current level, where it is close to the sound speed in a ``warm'' $HI$. In this case, a subsequent decrease of $M_{HI}$ down to modern values in the process of  disc evolution requires  that gas accretion have not compensated gas consumption for star formation -- at least in the last few billions years. In order for the relations $M_{HI} - V_{rot}D_{25}$ or $M_{HI} - V_{rot}R_0$ to remain linear, starformation efficiency SFE (star formation rate per gas mass unit) or its inverse value $\tau_{HI}$ (a gas consumption time) should be similar for galaxies with different masses and angular momenta.

Indeed, the available data are consistent with the approximately constancy of $\tau_{HI}$. Thus, for about 190 massive galaxies of $GASS$ survey with stellar masses $M_s>10^{10}M_\odot$ the average $\tau_{HI}$ was found to be $3\cdot 10^9$ years, regardless of galaxies mass (Schiminovich et al. 2010). Later, Wong et al. (2016) obtained  the mean SFE value, corresponding to $\tau_{HI}$ = $4.5\cdot 10^9$ years with the dispersion 0.3 dex, for a wide range of masses and rotational velocities of galaxies -- from dwarf to giant ones. Note that a consumption time $\tau_{HI}$ can be considered as a minimum time interval, passed since the gas mass in the discs was several times higher than at present.

It should be pointed, that within each galaxy $SFE$ may not be constant (it usually decreases with the radial distance), so it can be expected that surface density profile $\Sigma_{HI}(R)$ experiences evolutionary changes, depending on starformation rate, accretion rate and the gas losses at a given $R$. To describe the evolution of gas content in disc of galaxies it requires the construction of multi-parameter models, what is beyond the scope of this paper. The main thing is that if $\tau_{HI}$ keeps approximately constant for galaxies with different masses, then the relations discussed above will remain log-linear during the long period of time. Of course, the naturally expected scatter of $SFE$  for different galaxies must eventually ``erode'' them. According to Wong et al.  (2016), the observed $SFE$ scatter for galaxies is about 0.3 dex (i.e. a factor of 2) and, taking into account the measurement errors, the actual $SFE$ scatter can be even lower. It confirms that the linear relationships we consider will not blur out for a few billion years. Note however, that  there exist galaxies with abnormally high $HI$ mass on the top of corridor at the diagram ``$M_{HI} - V_{rot}D_{25}$'' marked by the dash lines (Figures 5). Unlike most of other galaxies, they seems to preserve the marginal stability condition $Q_g \approx Q_{c,g} \le 2$ to the present epoch.

\medskip
\c{\bf 5. Conclusions}
 \medskip

1. We found a systematic shift by about 0.2 dex between the $M_{HI}$ estimates for edge-on and for isolated galaxies with similar sizes or rotational velocities. Apparently, this discrepancy is due to the overestimation of correction for self-absorption of $HI$-line fluxes used by $HYPERLEDA$ database for edge-on galaxies.

2. For the two samples of late-type galaxies we confirm  the presence of the close relationships found earlier between total hydrogen mass $M_{HI}$  and specific disc angular momentum $J$, that we consider proportional to the rotational velocity product by optical diameter ($V_{rot}D_{25}$) or by radial disc scalelenght ($V_{rot}R_0$).

3. The discussed relationships $M_{HI}-V_{rot}D_{25}$ and $M_{HI}-V_{rot}R_0$ can be used for $HI$ content diagnostics, allowing to reveal the galaxies with abnormally high or abnormally low $HI$ mass compared with the isolated  late-type galaxies having similar kinematic characteristics. As an example, the abnormally $HI$-rich galaxies  taken from $HIghMass\, galaxy\, sample$ (Huang et al. 2014) have $M_{HI}$ which is systematically higher than the isolated galaxies with a similar size and rotational velocity, while the low-brightness galaxies follow the main sequence at the diagram $M_{HI}-V_{rot}R_0$.

4. The relations between $M_{HI}$ and $J$ can be explained assumibg that the Toomre' stability parameter $Q_g$ is approximate constant for a gas layer in a wide range of radial distances. However, the existing estimates of $Q_g$ for late-type galaxies are on average 2-3 times higher than the critical values $Q_{c,g}\approx $1-2 for marginal stability. Only for some $HI$-rich galaxies the observed gas mass $M_{HI}$ agrees with that expected for marginally stable gas layers.

5. We consider two possible explanations for the existence of close-to-linear relationship between $M_{HI}$ and specific angular momentum $J$: either critical value of Toomre' parameter is several times higher than the usually accepted value  $Q_{c,g}\sim 1-2$, enabling gas layers of galaxies to be close to marginally stable state presently, or this relationship has formed in the past, when a  mass of gas in galactic discs was several times higher than at present, so that gas layers were close to marginally stable state only at that early epoch. After that galaxies have slowly decreased their gas content during the evolution down to modern values. This explanation removes the conflict with observation data, which evidence that  $Q_g\leq Q_{c,g}$, being also in a good consistency with the observed slow decrease of $HI$ mass in the Universe over the past several Gyr (Neeleman et al. 2016). However, it excludes the models where gas inflow and gas losses are balanced i.e. it requires that in most galaxies gas consumption for starformation should not be compensated by gas accretion in the disc.

We acknowledge Russian Science Foundation support (project No. 14-22-00041).
We acknowledge the usage of the HyperLeda database (http://leda.univ-lyon1.fr).

\bigskip
\pagebreak
 
\pagebreak


{\tiny
\begin{longtable}{|c|c|c|c|c|c|c|c|c|}
\caption{Final relations}\label{Tabl1} \\
\hline
  & \multicolumn{4}{|c|}{\textbf{Edge-on}} & \multicolumn{4}{|c|}{\textbf{AMIGA}}\\
\hline
{\it 1} & {\it 2} & {\it 3} & {\it 4} & {\it 5} & {\it 2} & {\it 3} & {\it 4} & {\it 5}\\
\hline
$f(x)=kx+b$ & $k$ & $b$ & $MSE$ & $p$ & $k$ & $b$ & $MSE$ & $p$\\
\hline
\endhead

\hline
\endfoot
{\textbf{Objects in the sample}} & \multicolumn{4}{|c|}{\textbf{256}} & \multicolumn{4}{|c|}{\textbf{293}}\\
\hline
$M_{HI}=f(D_{25})$ & 1.43 & 7.90 & 0.079 & 0.80 & 1.59 & 7.50 & 0.076 & 0.87 \\
$M_{HI}=f(V_{rot})$ & 2.26 & 5.00 & 0.084 & 0.78 & 1.83 & 5.69 & 0.144 & 0.73 \\
$M_{HI}=f(V_{rot}D_{25})$ & 0.95 & 6.52 & 0.069 & 0.83 & 0.98 & 6.24 & 0.079 & 0.86 \\
\hline
{\textbf{Objects in the sample}} & \multicolumn{4}{|c|}{\textbf{256}} & \multicolumn{4}{|c|}{\textbf{71}}\\
\hline
$M_{HI}=f(R_0)$ & 1.45 & 9.02 & 0.092 & 0.76 & 1.49 & 8.85 & 0.071 & 0.83 \\
$M_{HI}=f(V_{rot}R_0)$ & 1.02 & 7.09 & 0.069 & 0.83 & 1.03 & 6.91 & 0.071 & 0.83 \\

\hline
\end{longtable}
Columns: (1)number (the relationship); (2), (3) linear regression parameters; (4) standard deviation; (5) correlation coefficient.
\pagebreak

\begin{longtable}{|c|c|c|c|}
\caption{Estimation of stability parameter $Q_g$ on radius $R_{25}$ }\label{Tabl2} \\
\hline {\it 1} & {\it 2}& {\it 3}& {\it 4} \\
\hline $NGC$ & $R_{25}$, kpc & $\sigma_{V}$, km/s & $Q_g$ \\\hline

\endhead
\hline
\endfoot
NGC925 & 14.3 & 8.8 & 1.8 \\
NGC2366 & 2.2 & 11.3 & 3.1 \\
NGC2403 & 7.4 & 8.4 & 2.9 \\
NGC2903 & 15.2 & 9.9 & 4.3 \\
NGC2976 & 3.8 & 9.6 & 13.8 \\
NGC3198 & 12.9 & 12.5 & 2.8 \\
IC2574 & 7.5 & 8.1 & 1.5 \\
NGC3621 & 9.4 & 10.0 & 3.3 \\
NGC4736 & 5.3 & 7.7 & 10.4\\
DDO154 & 1.2 & 8.7 & 2.4 \\
NGC5055 & 17.2 & 8.9 & 3.5 \\
NGC6946 & 9.8 & 7.7 & 4.6 \\
NGC7793 & 5.9 & 9.6 & 2.3 \\

\hline
\end{longtable}
Columns: (1) galaxy name; (2) isophote radius; (3) $HI$ velocity dispersion (according Ianjamasimanana et al. 2015); (4)  $Toomre$ stability parameter $Q_g$ on $R_{25}$.
\pagebreak

}

\begin{figure} [h!]
\includegraphics[width=19cm,keepaspectratio]{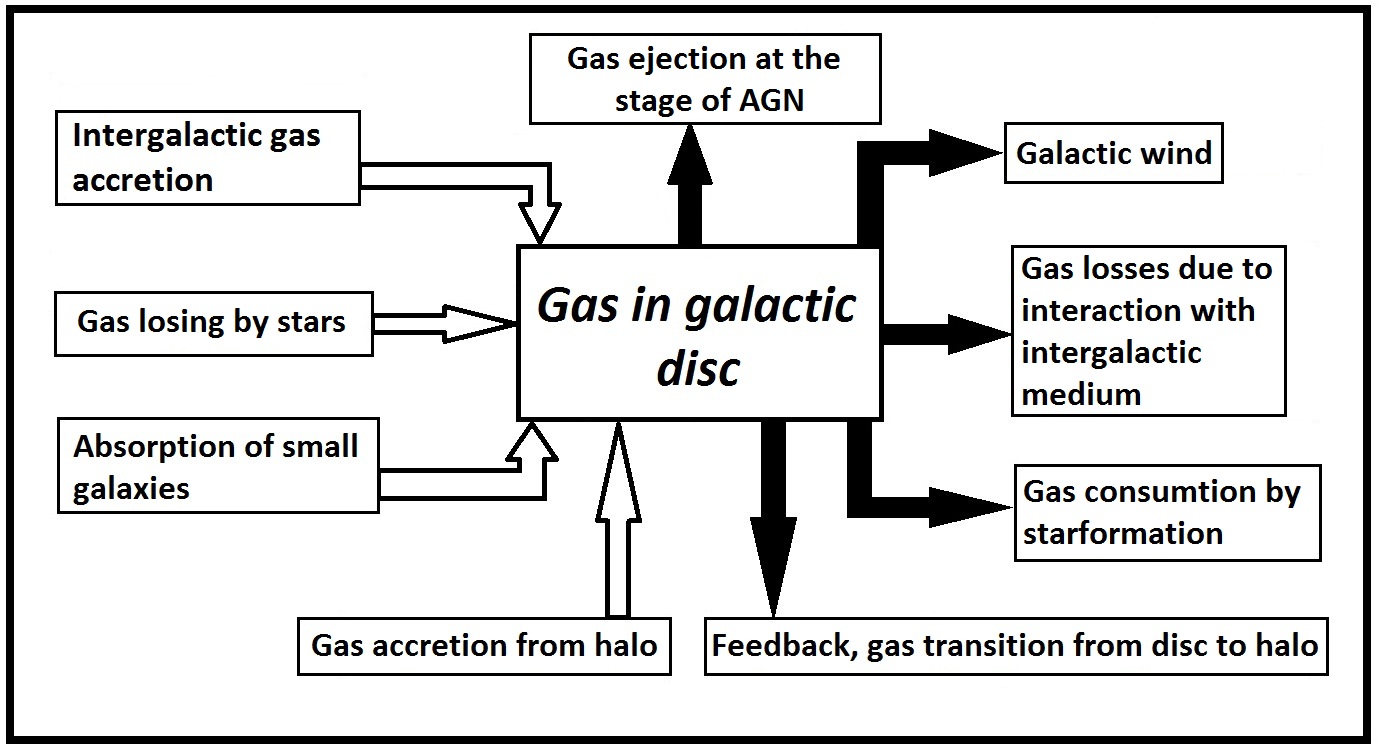}
\caption{Processes determining the evolution of gas in discs of galaxies.}
\label{ris:image1}
\end{figure}

\begin{figure} [h!]
\includegraphics[width=19cm,keepaspectratio]{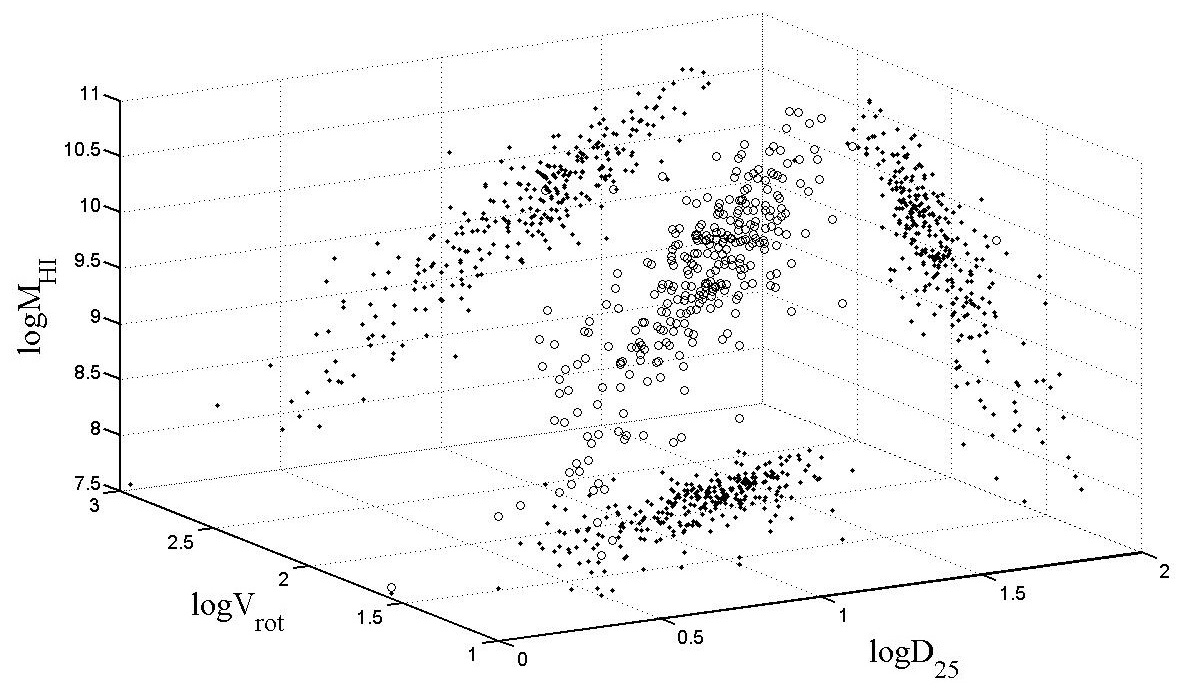}
\caption{The relationships between hydrogen mass, rotation speed and disc linear size for isolated late-type galaxies.}
\label{ris:image2}
\end{figure}

\begin{figure} [h!]
\includegraphics[width=19cm,keepaspectratio]{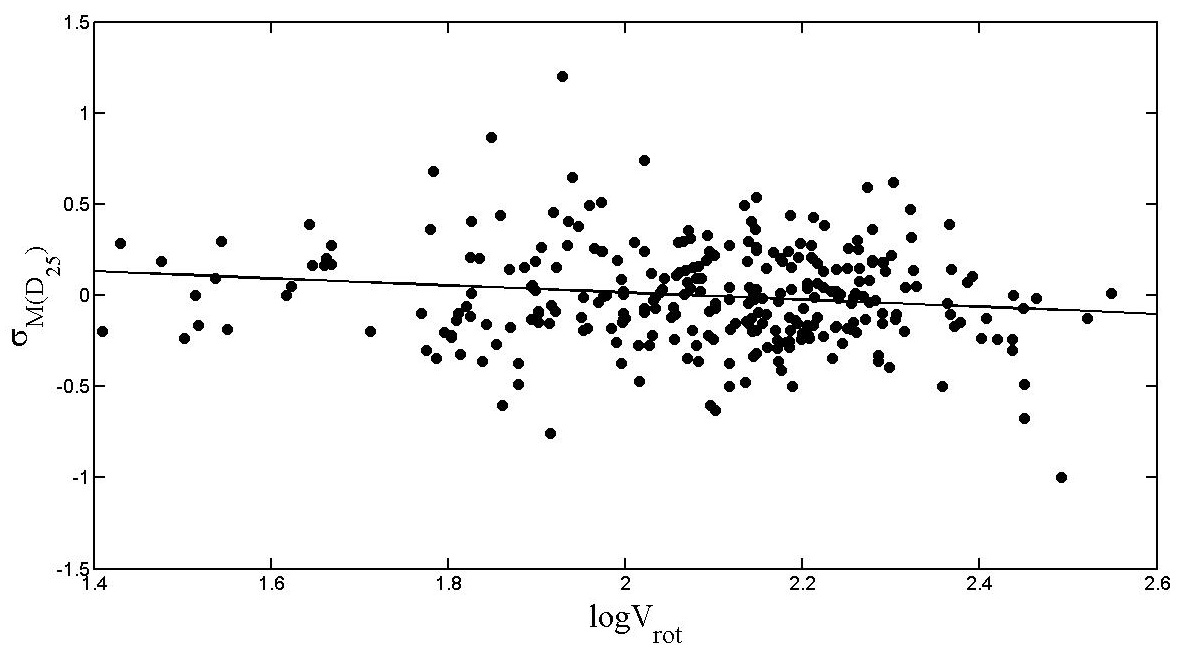}
\caption{The ratio between the galaxies rotation speed and deviations from the regression line at ``$M_{HI}-D_{25}$'' diagram.}
\label{ris:image3}
\end{figure}

\begin{figure}[h]
\begin{minipage}[h]{1\linewidth}
\center{\includegraphics[width=1.1\linewidth]{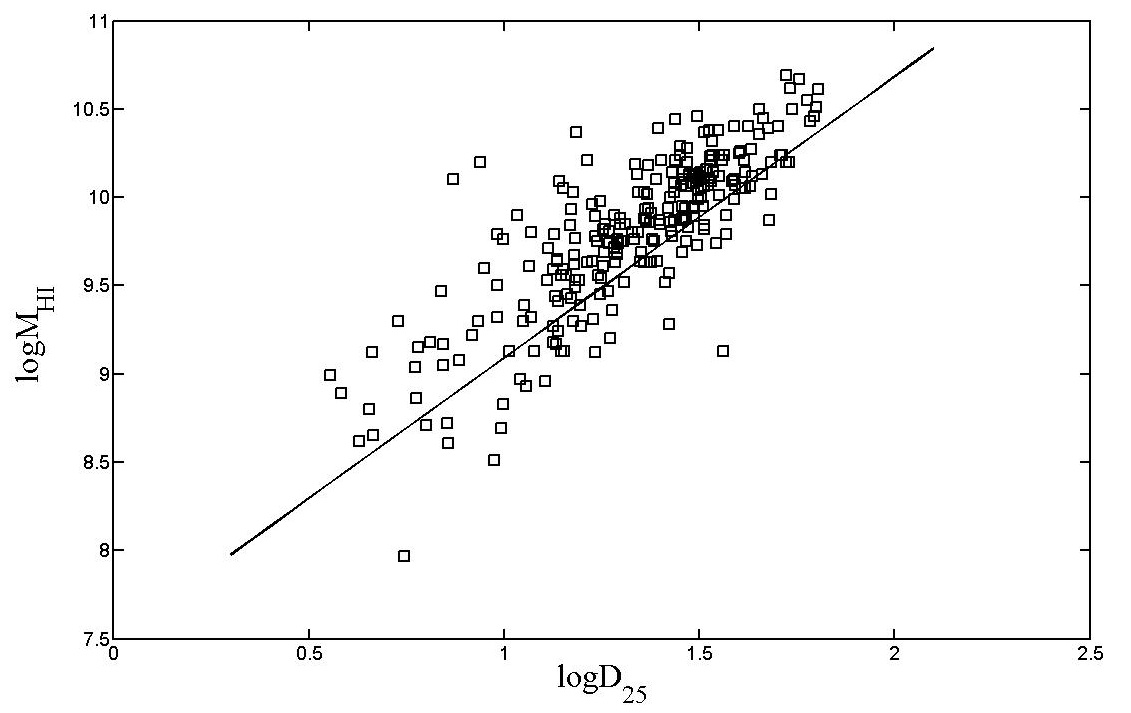} \\ a)}
\end{minipage}
\vfill
\begin{minipage}[h]{1\linewidth}
\center{\includegraphics[width=1.1\linewidth]{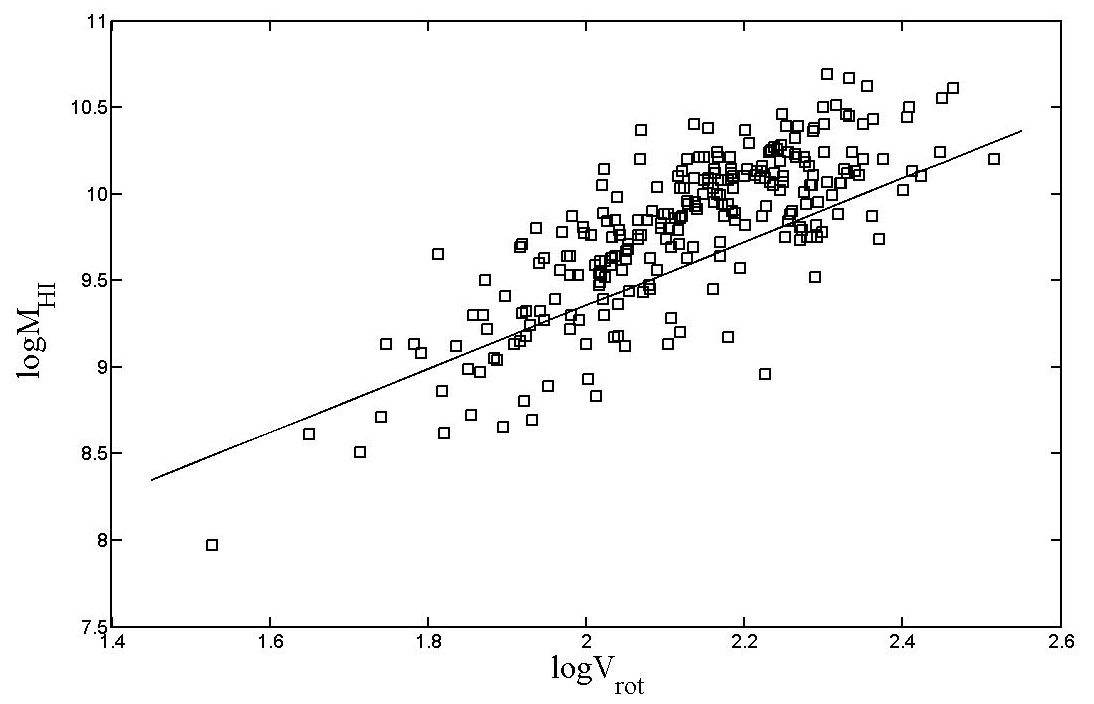} \\ b)}
\end{minipage}
\caption{Edge-on galaxies position at ``$M_{HI}-D_{25}$'' (a) and at ``$M_{HI}-V_{rot}$'' (b) diagrams. Straight line represents the regression for isolated galaxies.}
\label{ris:image4}
\end{figure}

\begin{figure}[h]
\begin{minipage}[h]{1\linewidth}
\center{\includegraphics[width=0.9\linewidth]{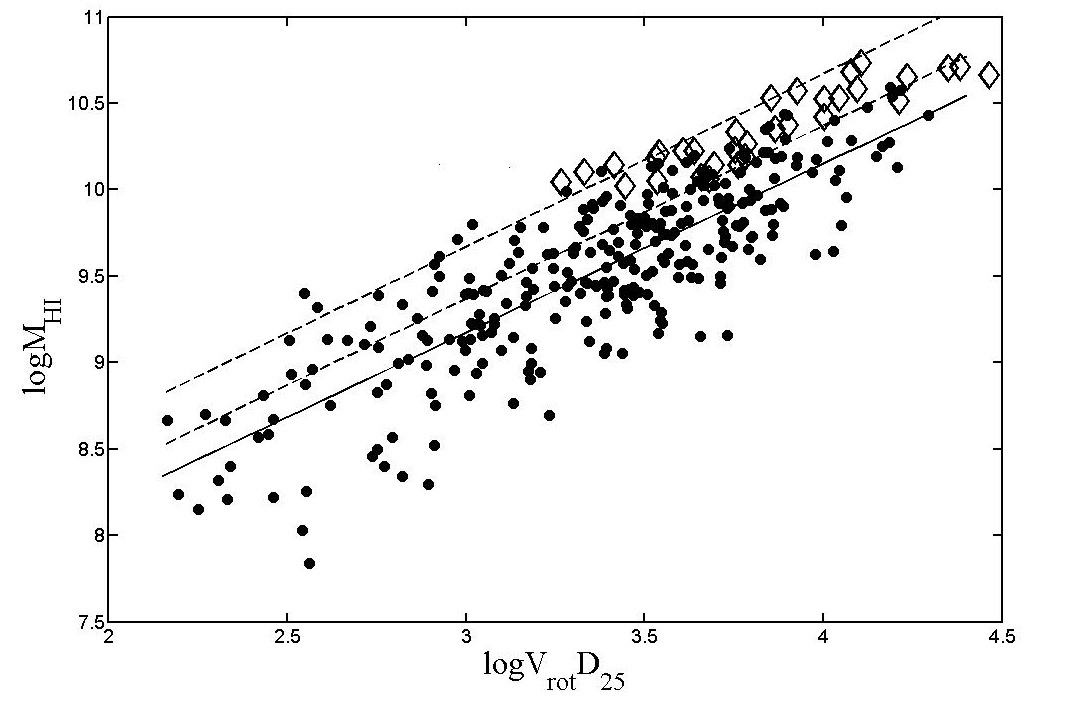} \\ a)}
\end{minipage}
\vfill
\begin{minipage}[h]{1\linewidth}
\center{\includegraphics[width=0.9\linewidth]{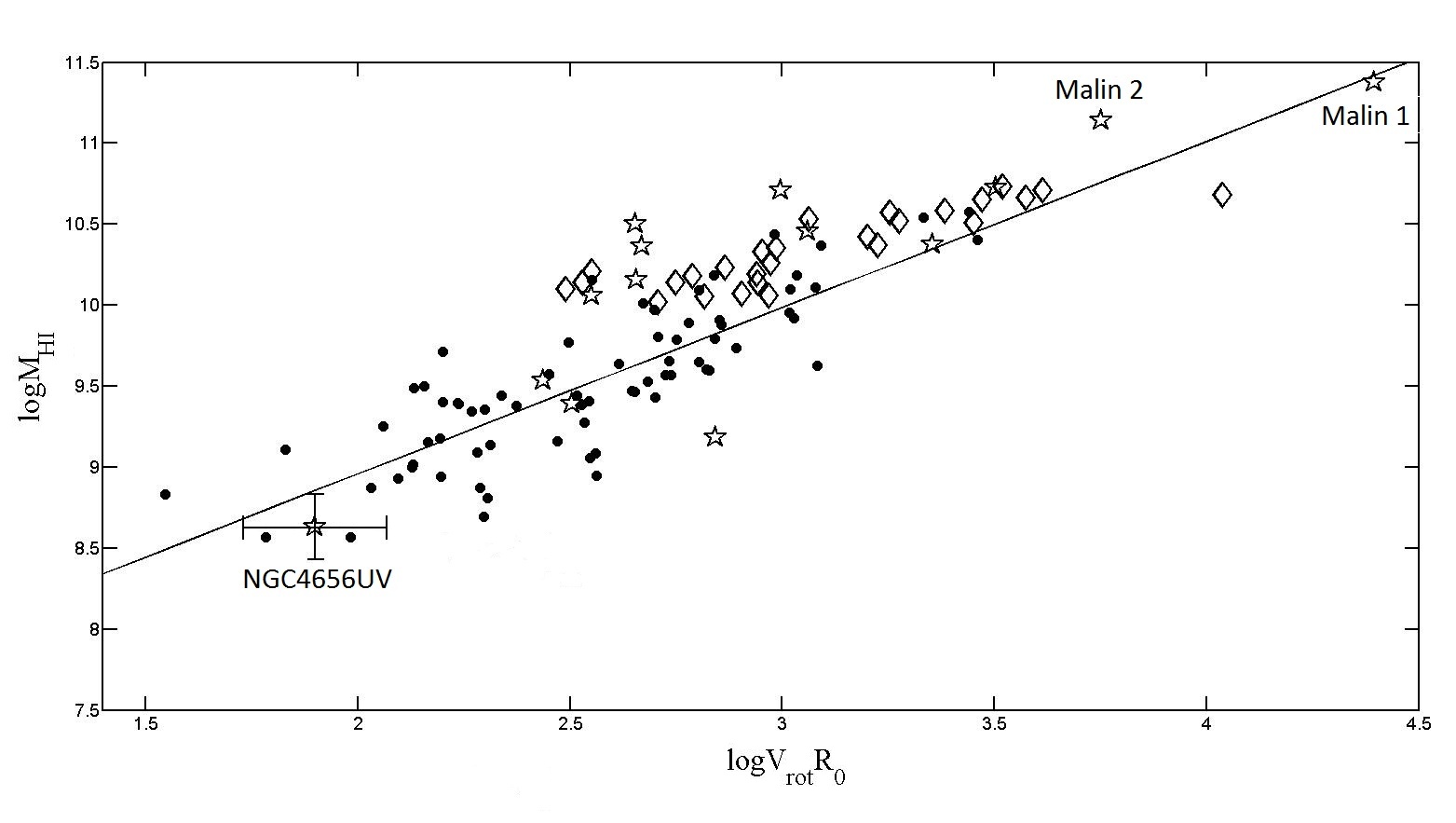} \\ b)}
\end{minipage}
\caption{Comparison of $HI$ mass $M_{HI}$ with kinematic parameters, proportional to specific galactic momentum: $V_{rot}D_{25}$ (a) and $V_{rot}R_0$ (b) for isolated galaxies (black points). The diamonds at both diagrams mark galaxies with abnormally high $HI$ content (Huang et al. 2014). Several LSB-galaxies which included $Malin \, 1,2$ and $NGC4656UV$ (asterisk) are presented at Fig. (b). Two parallel dash lines corresponds to the expected relations for marginally stable gas layer with flat rotation curves for the expected range of $K =\frac {c}{Q_g}$ = 10 km/s and 5 km/s (upper and lower lines respectively) (see the text). Solid straight line is linear regression line for isolated galaxies.}
\label{ris:image5}
\end{figure}


\medskip
\begin{thebibliography}{18}

\bibitem{1}
Abramova O.V., Zasov A.V., Astron. Rep., \textbf{55}, 202 (2011)

\bibitem{2}
Begum A., Chengalur J.N., Karachentsev I.D., Sharina M.E., MNRAS, \textbf{386}, 138 (2008)

\bibitem{3}
Bigiel F., Leroy A., Walter F., Blitz L., Brinks E., de Blok W.J.G., Madore B., AJ, \textbf{140}, 1194 (2010)

\bibitem{4}
Bigiel F., Blitz L., ApJ, \textbf{756}, 183 (2012)

\bibitem{5}
Boissier S., Gil de Paz A., Boselli A., Madore B.F., Buat V., Cortese L., Burgarella D., Munoz-Mateos J.C. et al., ApJSS, \textbf{173}, 524 (2007)

\bibitem{6}
Bothwell M.S., Wagg J., Cicone C., Maiolino R., Moller P., Aravena M., De Breuck C., Peng Y.  et al., MNRAS, \textbf{445}, 2599 (2014)

\bibitem{7}
Bradford J.D., Geha M.C., Blanton M.R., ApJ, \textbf{809}, 146 (2015)

\bibitem{8}
Brammer G.B., Whitaker K.E., van Dokkum P.G., Marchesini D., Labbe I., Franx M., Kriek M., Quadri R.F. et al., ApJL, \textbf{706}, L173 (2009)

\bibitem{9}
Bundy K., Scarlata C., Carollo C.M., Ellis R.S., Drory N., Hopkins Ph., Salvato M., Leauthaud A. et al., ApJ, \textbf{719}, 1969 (2010)

\bibitem{10}
Courtois H.M., Tully R.B., Fisher J.R., Bonhomme N., Zavodny M., Barnes A., ApJ, \textbf{138}, 1938 (2009)

\bibitem{11}
de Blok W.J.G., Walter F., Brinks E., Trachternach C., Oh S.-H., Kennicutt R.C., ApJ, \textbf{136}, 2648 (2008)

\bibitem{12}
de Grijs R., Peletier R.F., A\& A, \textbf{320}, L21 (1997)

\bibitem{13}
Elmegreen B.G., ApJ, \textbf{737}, 10 (2011)

\bibitem{14}
Elmegreen B.G., Hunter D.A., ApJ, \textbf{805}, 145 (2015)

\bibitem{15}
Evoli C., Salucci P., Lapi A., Danese L., ApJ, \textbf{743}, 45 (2011)

\bibitem{16}
Hall M., Courteau S., Dutton A.A., McDonald M., Zhu Y., MNRAS, \textbf{425}, 2741 (2012)

\bibitem{17}
Heidmann J., Heidmann N., de Vaucouleurs G., Mmras, \textbf{75}, 85 (1972)

\bibitem{18}
Huang Sh., Haynes M.P., Giovanelli R., Brinchmann J., ApJ, \textbf{756}, 113 (2012)

\bibitem{19}
Huang Sh., Haynes M.P., Giovanelli R., Hallenbeck G., Jones M.G., Adams E.A.K., Brinchmann J., Chengalur J.N. et al., ApJ, \textbf{793}, 40 (2014)

\bibitem{20}
Ianjamasimanana R., de Blok W.J.G., Walter F., Heald G.H., Caldu-Primo A., Jarrett Th.H. et al., AJ, \textbf{150}, 47 (2015)

\bibitem{21}
Karachentsev I.D., Karachentseva V.E., Kudrya Yu.N., Sharina M.E., Parnovskij S.L., Bulletin of the Special Astrophysical Observatory, \textbf{47}, 185 (1999)

\bibitem{22}
Karachentsev I.D., Karachentseva V.E., Huchtmeier W.K., Makarov D.I., ApJ, \textbf{127}, 2031 (2004)

\bibitem{23}
Karachentsev I.D., Makarov D.I., Kaisina E.I., AJ, \textbf{145}, 22 (2013)

\bibitem{24}
Karachentseva V.E., Lebedev V.S., Shcherbanovskij A.L., Catalogue of Isolated Galaxies. Bull. Inf. CDS, \textbf{30}, 125 (1986)

\bibitem{25}
Kennicutt R.C., ApJ, \textbf{344}, 685 (1989)

\bibitem{26}
Khoperskov A.V., Zasov A.V., Tyurina N.V., ARep, \textbf{47}, 357 (2003)

\bibitem{27}
Kim W.-T., Ostriker E.C., ApJ, \textbf{559}, 70 (2001)

\bibitem{28}
Kim W.-T., Ostriker E.C., ApJ, \textbf{660}, 1232 (2007)

\bibitem{29}
Lelli F., McGaugh S.S., Schombert J.M., AJ, \textbf{152}, 157 (2016)

\bibitem{30}
Leroy A.K., Walter F., Brinks E.,; Bigiel F., de Blok W.J.G., Madore B., Thornley M.D., ApJ, \textbf{136}, 2782 (2008)

\bibitem{31}
Lisenfeld U., Verdes-Montenegro L., Sulentic J., Leon S., Espada, D., Bergond G., Garcia E., Sabater J. et al., A\& A, \textbf{462}, 507 (2007)

\bibitem{32}
Lisenfeld U., Espada D., Verdes-Montenegro L., Kuno N., Leon S., Sabater J., Sato N., Sulentic J. et al., A\& A, \textbf{534}, 25 (2011)

\bibitem{33}
Makarov D., Prugniel Ph., Terekhova N., Courtois H., Vauglin I., A\& A, \textbf{570}, A13 (2014)

\bibitem{34}
Martin C., Kennicutt R., ApJ, \textbf{555}, 301 (2001)

\bibitem{35}
Martinsson Th.P.K., Verheijen M.A.W., Bershady M.A., Westfall K.B., Andersen D.R., Swaters R.A., A\& A, \textbf{585}, A99 (2016)

\bibitem{36}
Meurer G.R., Zheng Zh., de Blok W. J. G., MNRAS, \textbf{429}, 2537 (2013)

\bibitem{37}
Neeleman M., Prochaska J.X., Ribaudo J., Lehner N., Howk J.Ch., Rafelski M., Kanekar N., ApJ, \textbf{818}, 113 (2016)

\bibitem{38}
Obreschkow D., Glazebrook K., Kilborn V., Lutz K., ApJ, \textbf{824}, L26 (2016)

\bibitem{39}
Polyachenko V.L., Polyachenko E.V. Strel'Nikov A.V., AstL, \textbf{23}, 483 (1997)

\bibitem{40}
Quirk W.J., ApJ, \textbf{176}, L9 (1972)

\bibitem{41}
Romeo A.B., Wiegert J., MNRAS, \textbf{416}, 1191 (2011)

\bibitem{42}
Romeo A.B., Falstad N., MNRAS, \textbf{433}, 1389 (2013)

\bibitem{43}
Russel D.G., ApJ, \textbf{565}, 681 (2002)

\bibitem{44}
Sachdeva S., Saha K., ApJL, \textbf{820}, 4 (2016)

\bibitem{45}
Safonova E.S., ARep, \textbf{55}, 1016 (1997)

\bibitem{46}
Schechtman-Rook A., Hess K.M., ApJ, \textbf{750}, 171 (2012)

\bibitem{47}
Schiminovich D., Catinella B., Kauffmann G., Fabello S., Wang J., Hummels C., Lemonias J., Moran S.M. et al., MNRAS, \textbf{408}, 919 (2010)

\bibitem{48}
Solanes J.M., Giovanelli R. and Haynes M.P., ApJ, \textbf{461}, 609 (1996)

\bibitem{49}
Stinson G.S., Dutton A.A., Wang L., Maccio A.V., Herpich J., Bradford J.D., Quinn T.R., Wadsley J. et al., MNRAS, \textbf{454}, 1105 (2015)

\bibitem{50}
Swaters R.A., van Albada T.S., van der Hulst J.M., Sancisi R., A\& A, \textbf{390}, 829 (2002)

\bibitem{51}
Toribio M.C., Solanes J.M., Giovanelli R., Haynes M.P., Martin A.M., ApJ, \textbf{732}, 93 (2011)

\bibitem{52}
Vulcani B., Poggianti B.M., Fritz J., Fasano G., Moretti A., Calvi R., Paccagnella A., ApJ, \textbf{798}, 14 (2015)

\bibitem{53}
Walter F., Brinks E., de Blok W.J.G., Bigiel F., Kennicutt R.C., Thornley M.D., Leroy A., ApJ, \textbf{136}, 2563 (2008)

\bibitem{54}
Wang J., Koribalski B.S., Serra P., van der Hulst Th., Roychowdhury S., Kamphuis P., Chengalur J.N., MNRAS, \textbf{460}, 2143 (2016)

\bibitem{55}
Westfall K.B., Andersen D.R., Bershady M.A., Martinsson Th.P.K., Swaters R.A., Verheijen M.A.W., ApJ, \textbf{785}, 43 (2014)

\bibitem{56}
Wong O.I., Meurer G.R., Zheng Z., Heckman T.M., Thilker D.A., Zwaan M.A., MNRAS, \textbf{460}, 1106 (2016)

\bibitem{57}
Yim K., Wong T., Howk J.Ch., van der Hulst J.M., ApJ, \textbf{141}, 48 (2011)

\bibitem{58}
Yim K., Wong T., Xue R., Rand R.J., Rosolowsky E., van der Hulst J.M., Benjamin R., Murphy E.J., ApJ, \textbf{148}, 127 (2014)

\bibitem{59}
Zasov A.V., Simakov S.G., Astrophysics, \textbf{29}, 518 (1988)

\bibitem{60}
Zasov A.V., Rubtsova T.V., Soviet AstL, \textbf{15}, 51 (1989)

\bibitem{61}
Zasov A.V., Sulentic J.W., ApJ, \textbf{430}, 179 (1994)

\bibitem{62}
Zasov A.V., Smirnova A.A., AstL, \textbf{31}, 160 (2005)

\bibitem{63}
Zasov A.V., Terekhova N.A., AstL, \textbf{39}, 291 (2013)

\bibitem{64}
Zasov A.V., Saburova A.S., Egorov O.V., Uklei R.I., \textbf{1705.03523}, accepted for publication in MNRAS (2017)

\end{thebibliography}
\end{document}